\documentstyle[twocolumn,floats,ifthen,aps,prb]{revtex} 

\begin{document}
\draft
\title{Shot noise suppression at 1D hopping }
\author{Alexander N. Korotkov and Konstantin K. Likharev}
\address{Department of Physics and Astronomy, 
State University of New York, Stony Brook, New York 11794-3800 }

\date{\today}

\maketitle

\begin{abstract}
We have carried out a preliminary analysis of shot noise at hopping, 
focusing on uniform 1D arrays of sites separated by $N$ tunnel 
barriers. The results show that at low temperatures the low-frequency density
of the shot noise varies from $1/N$ to 1 of the Schottky value, depending on 
the geometry, electron density, and Coulomb interaction strength. 
An interesting feature is $\omega^{-1/3}$ dependence of the current 
spectral density at intermediate frequencies, which reflects self-similarity 
of the fluctuations at different size scales.
\end{abstract}

\pacs{}


\narrowtext

\section{Introduction}

Nonequilibrium fluctuations in mesoscopic systems can present additional
information that is not reflected in their dc transport characteristics.
This is one of the reasons why ``shot noise'' (i.e.\ nonequilibrium 
fluctuations of current with constant or nearly-constant spectral density 
at low frequencies) has attracted so much attention in mesoscopics during 
the last decade - see, e.g., Refs. \cite{Kogan,deJong,Blanter}. 

An additional motivation for the present paper was provided by the 
observation \cite{Matsuoka} that the 
smallness of the shot noise is a necessary condition for quasi-continuous
electron transfer. More exactly, for an external observer a conductor
provides effectively Ohmic (quasi-continuous) conduction only if the
so-called Fano factor 
\begin{equation}
F\equiv S_{I}(0)/2e\langle I\rangle  \label{FanoDef}
\end{equation}
(where $S_{I}(0)$ is the low-frequency density of current fluctuations, and 
$\langle I\rangle $ is the average current) is much lower than 1. If
simultaneously the resistance of such a sample is sufficiently high, and its
stray capacitance is low,
\begin{equation}
R \gg \hbar /e^{2}, \, C\ll e^{2}/T, \, \label{C<<e2/T}
\end{equation}
it may be used for resistive coupling in single-electron devices. Since
using resistively-coupled devices is one of the very few options available 
to avoid the 
forbidding problem of random background charge in single-electronics (see, 
e.g., Refs. \cite{LikhSETrev,KorSETrev}), the search for systems with 
quasi-continuous conduction is important for possible future applications
of single-electron devices in integrated circuits.

Shot noise has been extensively analyzed for metallic conduction (in 
both ballistic \cite{Lesovik,Martin,Naveh-bal} and diffusive \cite
{Nagaev92,Beenakker,Nagaev95,Naveh9799,Naveh98} limits) and for 
single-electron tunneling. \cite{Kor-92,Kor-94,Hershfield,Galperin}
Unfortunately, metallic conductors can satisfy the condition $F\ll 1$ only
if they are much longer than the electron-phonon interaction length. 
\cite{Nagaev95,Naveh98} As a result Eqs.\ (\ref{C<<e2/T}) can be practically 
met only at very low temperatures - see, e.g., experiment \cite{Wei}. The 
same is true for single-electron circuits (like 1D or 2D arrays) with their 
relatively large islands \cite{LikhSETrev,KorSETrev}.

Much higher resistance $R$ at small sample length (and hence small $C$) is
typical for hopping conductors - see, e.g., Refs.\ \cite{Mott,Shklovskii}.
Naively, one might think that since the hopping transport is due to discrete
single electron tunneling events (``hops''), the shot noise should be close 
to the Schottky value ($S_{I}(0)=2e\langle I\rangle$, i.e., $F=1$). However, 
this argument is obviously not true, since it could also be applied to a 1D 
series array  of $N$ tunnel junctions. A simple ``circuit'' theory 
\cite{Ziel,Landauer} (see also Appendix) shows that for such an array 
the Fano factor can be very small: 
	\begin{equation}
F\sim 1/N \ll 1.  \label{F<<1} 
	\end{equation}
The physical reason for this fact is that the noise originating from each 
junction is strongly shunted by the junction resistance, which is much
smaller than the total resistance of other junctions. 
          
Thus, there is hope of having the shot noise at hopping 
suppressed well below the Schottky value as well. However, the real picture 
of hopping is complex, and the noise may be much higher than the simple 
estimate given above. For example, mutual correlation of the hopping events, 
exponentially broad distribution of their rates due to sample randomness,
and the percolative character of transport paths in 2D and 3D 
cases \cite{Mott,Shklovskii} may all be important factors. Until recently, 
the situation was virtually unexplored:
the few publications on the theory of noise in hopping we are aware of (see,
e.g., Ref.\ \cite{Kogan-hop} and references therein) concentrate on $1/f$
noise rather than on the broadband shot noise.\cite{shot} 
 We are also unaware of any
experimental studies of noise at hopping at frequencies high enough to avoid 
$1/f$ noise dominance.

The goal of this paper is to develop an initial picture of
shot noise at hopping. We will focus on the 1D case, and assume
uniformity of hopping conditions between all the sites.  
(A brief analysis of nonuniform systems and higher dimensions is 
given in Discussion.) In principle, 1D hopping may be implemented
experimentally using a linear array of quantum dots between two external
electrodes (Fig.\ \ref{schematic}a). Besides this geometry, we will 
also consider a somewhat artificial model of hopping on a ring (Fig.\ 
\ref{schematic}b), at least because problems with periodic boundary 
conditions are traditional in theoretical studies of hopping. Besides 
that, since such models automatically conserve the total electron number, 
they may crudely mimic ``open'' models (Fig.\ \ref{schematic}a) with 
considerable Coulomb interaction without its explicit account.

Throughout our analysis we will assume that the electron states localized at
each site are non-degenerate, so that each site may be occupied with just
one electron, or none. This model can be viewed as a special case of 
the ``orthodox'' theory of single-electron tunneling \cite{AvLikh91}
when the background charge of each island is close to $-e/2$, so that 
energies of two charge states ($n$ = 0 and $n$ = 1) are close to each other 
while other states are far beyond the available energy range. 
So, the well-developed theory of noise based either on Fokker-Planck 
\cite{Kor-92,Kor-94,Hershfield,Galperin} or Langevin \cite{Kor-98} approach 
can be directly applied to any hopping structure with arbitrary 
electron-electron interaction. However, these approaches involve taking 
into account an exponentially large number of charge configurations, thus 
limiting practical calculations to relatively small structures, 
$N\lesssim 20$. 
        This is why for the numerical results we have used the Monte-Carlo 
approach, similar to that used for simulations of transport \cite{Likh-Bakh} 
and noise \cite{Kor-94,Matsuoka} at single-electron tunneling, with the 
corresponding restriction of the site state number. 

        It is instructive to compare the results for the shot noise at
1D hopping and at tunneling in 1D array of tunnel junctions. Some formulas 
necessary for this comparison are derived in Appendix for the cases of 
negligible and small charge discreteness effects.

\section{Some general relations}

In the hopping limit, where quantum interference between states before and
after each hopping event is neglected because of the inelastic nature of 
electron transport, \cite{Mott,Shklovskii} site occupation numbers may be 
considered as random classical variables. If we are not interested in 
extremely high frequencies (when the finite photon energy becomes important),
current $I_{i}(t)$ flowing between the $(i-1)$th and $i$th site 
may be considered as a sum of infinitely short pulses: 
\begin{equation}
I_{i}(t)=I_{i}^{+}(t)-I_{i}^{-}(t),\text{ \ }I_{i}^{\pm
}(t)=\sum_{t_{k}}e\delta (t-t_{k}^{\pm }),  \label{Ii(t)} 
\end{equation}
where $t_{k}^{+}$ ($t_{k}^{-}$) is the time of $k$th hop in the positive
(negative) direction between the sites. In the ``open boundary'' problem
(Fig.\ \ref{schematic}a), with a fixed voltage across the sample, 
we may also consider
currents $I(t)$ flowing in external electrodes.\cite{Matsuoka,Likh-Bakh} 
These currents contain contributions not only from the hops to and from the
electrodes, but also the polarization charge changes (displacement current
contribution) due to hops between internal sites:
	\begin{equation}
I(t)=\sum_{i=1}^{N}\lambda _{i}I_{i}(t),\,\,\,\sum_{i}\lambda _{i}=1,
	\label{Ie(t)}\end{equation}
where the factors $\lambda _{i}$ depend on the structure geometry and can be
expressed via its electrostatic matrix. 
(In general, these coefficients are 
different for the left and right electrodes.) In the simplest case of a 1D 
array between two infinite parallel metallic plates, $\lambda _{i}=a_{i}/L$,  
$L\equiv \sum_{i}a_{i}$, where $a_i$ is transport direction component of 
$i$th hop vector. In this work, we will use this formula, with 
$a_{i}=L/N=$ const (i.e., $\lambda _{i}=1/N$), even for the ring geometry
(Fig.\ \ref{schematic}b), though this model does not have any electrodes. 
This assumption is not critical for the Fano factor which does not
depend on $\lambda_i$, since at low frequencies the 
spectral densities of all currents $I_{i}$ and $I$ coincide. 
(A simple proof of this statement may be obtained from the spectral 
density definition 
	\begin{equation}
S_{I}(\omega )=\lim_{\tau \rightarrow \infty }\frac{2}{\tau }\langle
|\int_{0}^{\tau }I(t) \, e^{\imath\omega t}dt|^{2}\rangle  
	\label{Sdef} \end{equation}
in the limit $\omega \rightarrow 0$, using the condition that the charge cannot
accumulate indefinitely inside the array). 
In the opposite limit of high frequencies (much
higher than the average tunneling rate, though still much lower than the
reciprocal ``time of tunneling'', which is considered infinitely 
short in our theory), the spectral densities of currents $I_{i}$ and $I$ 
are typically different, and obey a simple formula. In fact, in the 
high-frequency
limit all tunnel events are effectively uncorrelated and the phases of
factors $\exp (\imath \omega t_{k}^{\pm })$ in Eq.\ (\ref{Sdef}) are random.
From this, we obtain 
	\begin{equation}
\, S_{I_{i}}(\infty )=2e(\langle I_{i}^{+}\rangle +\langle I_{i}^{-}\rangle ),
\,\, S_{I}(\infty )=\sum_{i}\lambda _{i}^{2}S_{I_{i}}(\infty ) . 
	\label{Sinf}\end{equation}
It is easy to see that for the current through one barrier $S_{I_{i}}
(\infty )/2e\langle I\rangle \geq 1$, while for the external current 
$S_{I}(\infty )/2e\langle I\rangle \geq 1/N$. 
We will mostly be interested in the readily measurable quantity 
$S_{I}(\omega )$ and its low frequency value $S_I(0)$. 

        For the numerical (Monte-Carlo) calculations of the spectral density 
we have directly used\cite{Kor-94} Eq.\ (\ref{Sdef}). The time period 
$\tau$ is chosen to be 
sufficiently long and the averaging is done over many such time periods. 
In practical calculations, it is 
important to keep the product $\omega \tau /2\pi $ integer in order 
to avoid numerical inaccuracy at low frequencies, and it is convenient 
to calculate simultaneously the spectral density at several overtones 
of certain basic (low) frequency.  
For several figures we have also used the newly developed method for
the calculation of spectral density, which
gives much faster convergence; this method will be described elsewhere.

\section{Circular array}

\subsection{The model}

We start with the auxiliary problem of hopping 
of a fixed number ($M$) of electrons on a uniform ring of $N>M$ sites. The
electron may hop to either of the neighboring sites, i.e.\ either 
clockwise (with a probability rate of $\Gamma ^{+}$) or 
counterclockwise (with rate $\Gamma^{-}<\Gamma^{+}$), but only if the
accepting site is empty. 
 The rates $\Gamma ^{\pm }$ should satisfy the
Gibbs relation 
\begin{equation}
\Gamma ^{-}/\Gamma ^{+}=\exp (-W/T),
\end{equation}
where $W$ is the energy difference between the neighboring
sites. (Due to the circular geometry, a conceptually sound, though 
impractical, way to create this difference is to increase the magnetic flux
through the ring area at a constant rate. However, we consider the 
circular array mostly as a simplification of the realistic linear array.) 

        For the comparison of the current noise with the Nyquist formula
we will need the total resistance of the ring which is naturally defined 
as 
        \begin{equation}
        R_\Sigma =V/\langle I\rangle,  
        \label{Rsum}\end{equation}
where the total ``voltage'' $V$ is defined as $NW/e$ (the dependence
of the tunneling rate on $W$ can be arbitrary).  

In the final part of our analysis we will include the particle interaction
following the unscreened Coulomb law, so that the potential energy of the
system is 
        \begin{eqnarray} 
U\{{\bf r}_{1},{\bf r}_{2},..,.{\bf r}_{M}\} && 
=e^{2}\sum_{i<j}\frac{1}{\left| 
{\bf r}_{i}-{\bf r}_{j}\right| }
\nonumber \\
&& =\frac{e^{2}}{a}\sum_{i<j}\frac{\pi }{N\sin
\pi \frac{\left| n_{i}-n_{j}\right| }{N}},  
        \label{Coulomb}\end{eqnarray}
where $n_{i}$ is the site occupied by the $i$th electron. The interaction is
included into the model by adding the corresponding change of $U$ at a hop to
that $(\pm W)$ describing the external field. Since in this case the 
tunneling rates are no longer constant, we will need to specify an 
explicit relation $\Gamma ^{\pm }(W)$. In this case we will assume  
\begin{equation}
\Gamma \equiv \Gamma ^{+}-\Gamma ^{-}=W/eR_{0}, 
\end{equation}
        where $R_0$ gives the scale of the effective resistance of a tunnel
barrier between adjacent sites.

\subsection{Single particle limit} 

        Let us assume $M=1$. \cite{holes} Then the current $I(t)$ consists 
of uncorrelated pulses, each transferring 
the charge $\pm e/N$, with rates  $\Gamma ^{+}$ and 
$\Gamma ^{-}$, respectively. This is equivalent to the conventional case 
of one tunnel junction with the electron charge substituted by $e/N$, 
hence   
        \begin{equation}
\langle I\rangle =e(\Gamma^+ -\Gamma^-)/N, 
        \label{single1}\end{equation}
        and the spectral density is frequency-independent,\cite{Kogan}  
$S_{I}(\omega )=S_I(0)$, with 
\begin{equation}
S_I(0)=\frac{2 e^{2}}{N^2}\,(\Gamma ^{+}+\Gamma ^{-}) 
=2e\langle I\rangle \,
\frac{1}{N}\,\coth \frac{W}{2T}\, .  
        \label{Ssmall} \end{equation} 

        Figure \ref{coth} shows the corresponding Fano factor 
$F=N^{-1}\coth (W/2T)$, 
as a function of $W$. In thermodynamic equilibrium, $W=0$, 
the noise satisfies the Nyquist formula, $S_{I}=4T/R_{\Sigma }$,
which remains valid while $W \ll T$. At $W\rightarrow 0$ the Fano factor 
tends to infinity because 
the average current vanishes while the equilibrium thermodynamic
fluctuations still produce a finite current noise.
        For $N\gg 1$ the Fano factor crosses unity at 
$W\simeq W_{c}=2T/N \ll T$. 
Let us emphasize that since $W_{c} \ll T$, the noise at this
crossover is still due to thermodynamically equilibrium fluctuations. 
Finally, if the applied field is high ($W\gg T$), the Fano factor is low: 
        \begin{equation}
F=1/N.  
        \label{F=1/N}\end{equation}
        Thus, as a matter of principle the shot noise suppression at hopping 
may be really very strong (proportional to the array length, just as in 
tunnel junction arrays).  Now let us 
examine how this suppression is affected by various factors.

\subsection{Low temperature, no interaction}

At $T\ll W$ (i.e., $\Gamma \approx \Gamma ^{+}\gg \Gamma ^{-}$) and in 
the absence 
of Coulomb interactions ($e^{2}/a\ll W$), but for arbitrary electron density 
$\rho \equiv M/N$ our model is reduced to the so-called Asymmetric Simple
Exclusion Process (ASEP) model which has been extensively studied during 
the past few years -- for a review see Ref. \cite{Derrida}. 
Within this model, all $N!/M!(N-M)!$ possible charge configurations of the 
system have equal probability for the arbitrary $N$ and 
$M$. \cite{Derrida} From this fact, the average current is readily 
calculated to equal 
        \begin{equation}
\langle I\rangle =e\Gamma \, \frac{M}{N} \, \frac{N-M}{N-1},  
        \label{Icircle}\end{equation} 
so that for a large system ($N,M\rightarrow \infty $)
        \begin{equation}
\langle I\rangle =e\Gamma \rho (1-\rho ).  
        \label{<I>}\end{equation}
 Notice that these expressions (as well as those below)
are obviously symmetric over the transformation $\rho \leftrightarrow 
1-\rho $ which interchanges electrons and holes. 
        From  Eq.\  (\ref{<I>}), the maximum value $\langle 
I\rangle_{max} = e\Gamma /4$ of 
dc current is achieved at $\rho =1/2$, which is a trade-off 
between increasing concentration $\rho$ and decreasing 
average velocity $\Gamma (1-\rho )$ of each electron
(in hops per unit time) because of other electrons blocking 
its hops.  

        Equation (\ref{<I>}) is exactly the result which could be 
anticipated in the complete absence of correlation between the hops. 
However, in fact these correlations {\it do} exist, as revealed, for
example, by the spectral density of the current.
        Figure \ref{circle0.3} shows the result of numerical 
calculation of $S_I(\omega )$ using the Monte-Carlo approach
for two concentrations, $\rho= 0.3$ and $\rho =0.5$, and several 
values of the array length $N$. 
The frequency dependence of the spectral density is obviously not flat
as it would be in an uncorrelated case. With increasing $N$ the spectral
density decreases and forms three distinct regions as a function of
frequency: low and high frequency saturation regions and almost 
power-law decay in between. 

        At high frequencies, in accordance with Eq.\ (\ref{Sinf}) 
        \begin{equation}
S_{I_{i}}(\infty )=2e\langle I\rangle ,\,\,\,S_{I}(\infty )=2e\langle
I\rangle /N,  
        \label{largeomega}\end{equation} 
the suppression of the external current fluctuations is maximal. 
Notice that the frequency $\omega_{h}$ of the crossover to this limit 
apparently does not depend on $N$, while the low-frequency 
crossover occurs at frequency $\omega_{l}$ which decreases with $N$ 
crudely as $\omega_{l} \propto N^{-3/2}$. (In tunnel junction arrays, 
$\omega_{l}$ scales as $N^{-2}$ -- see Appendix). 
        The zero frequency limit has been followed analytically 
\cite{Derrida} giving the following Fano factor:
        \begin{equation}
F=\frac{\pi ^{1/2}}{2}\left[ \frac{\rho (1-\rho )}{N}\right] ^{1/2},
\,\, \text{for } N,M\rightarrow \infty  
        \label{Fcircle}\end{equation}
(an analytical formula is also available  \cite
{Derrida} for arbitrary $N$ and $M$).
        Figure \ref{circle0.3} shows that at large $N$ the frequency 
dependence of the current spectral density in the intermediate frequency 
range approaches power law: $S_I(\omega )\propto \omega^{-1/3}$.

This dependence may be interpreted as a consequence of the self-similarity
of the fluctuations \cite{SOC}  which occur at any site number scale $L$, 
within the 
interval $1\ll L\ll N$. In order to explain the $\omega^{-1/3}$ scaling, 
let us assume that Eq.\ (\ref{<I>}) is applicable to 
long-wave density perturbations in our system 
and introduce two velocities (measured in sites per second) 
of their propagation. 

        The first of them, the sound (``shock'' \cite{Derrida}) 
velocity  
        \begin{equation}
v_{s}=(1-2\rho )\Gamma ,  
        \label{vs}\end{equation} 
can be found from the obvious continuity equation $\partial (e\rho) /
\partial t=-\partial I/\partial x$, where $I$ and $\rho $ 
are understood in the 
sense of ``local'' averages over $1\ll \delta N\ll N$ sites and 
$x$ is the site number considered as a continuous coordinate. Since 
these averages are related by Eq.\ (\ref{<I>}), for small deviations from 
equilibrium we get $\partial \rho /\partial t$ $=-(1-2\rho )\Gamma
\,\partial \rho /\partial x$, i.e.\ an equation describing linear waves
propagating with the speed given by Eq.\ (\ref{vs}). Notice that the sound  
velocity vanishes at half-filling, $\rho =1/2$, 
and is negative beyond this point. 

        In the circular array all density fluctuations move with the 
same sound velocity, so the fluctuation profile does not evolve in time 
and thus overall rotation does not affect the noise of current $I$ 
($v_s$ will, however, be important later for the analysis of the linear 
array). To study the relaxation of density fluctuations we 
need to consider the deviations of $v_s$,  
        \begin{equation}
\delta v \simeq -\Gamma \delta \rho . 
        \label{dv}\end{equation}
(Including the factor 2 following from Eq.\ (\ref{vs}) would be an overestimate
of our accuracy, since such nonlinear velocity can be defined 
in various ways leading to different numerical coefficients). 

        To calculate fluctuations $I(t)$ at a frequency $\omega \ll 
\Gamma$, we can integrate Eq.\ (\ref{<I>}) over the whole circle 
taking into account local density fluctuations $\delta \rho$. Since
we have assumed uniform $\lambda_i$ in Eq.\ (\ref{Ie(t)}) and the total
number of electrons does not fluctuate, $\int \rho (x) dx =M$, 
the contribution from the linear term  $\delta I  =
e\Gamma \delta \rho$ vanishes. 
         However, the current 
fluctuations do appear in the next, quadratic term of Eq.\ (\ref{<I>}): 
$\delta I = -e\Gamma (\delta \rho )^{2}$, which 
describes the ``rectification'' of density fluctuations. 

        The density fluctuations at the size scale $L$ ($1 \ll L  
\ll N$) are described by the binomial distribution, giving the variance 
$\langle (\delta \rho )^2 \rangle = \rho (1-\rho )/L$. Hence, 
the typical relaxation bandwidth of these fluctuations (in the frame rotating
with velocity $v_s$) is $\omega_L \simeq 
|\delta v|/ L \simeq \Gamma [\rho(1-\rho )]^{1/2}L^{-3/2}$, 
and the corresponding spectral density is $S_\rho (\omega_L)
\simeq (\delta \rho )^2/\omega_L \simeq [\rho (1-\rho ) L]^{1/2}/\Gamma $.
        According to the standard theory of noise rectification (see, e.g., 
\cite{Rytov}), $S_I(\omega_L)$ can be estimated as 
$(N/L)e^{2}\Gamma ^{2}[S_{\rho }(\omega_L)]^{2} \omega_L $,  where the 
first factor accounts for $N/L$ virtually independent fluctuating regions.
Combining these estimates and eliminating $L$ (as a function of $\omega_L$),
we finally obtain 
        \begin{equation}
\frac{S_{I}(\omega )}{2e\langle I\rangle } \simeq C\, 
\frac{(\omega /2\pi \Gamma )^{-1/3}}{N} 
\,[\rho (1-\rho )]^{2/3},  
        \label{S(w)}\end{equation}
where the numerical factor $C$ can be found by comparison with the 
Monte-Carlo results (Figs.\ \ref{circle0.3} and \ref{circle2}), 
giving a value between 1.1 and 1.2. 

        Notice that Eq.\ (\ref{S(w)}) is accurate only if both $N$ and $M$
are sufficiently large. Figure \ref{circle2} shows $S_I(\omega )$ normalized
by the value $2e\langle I\rangle (\omega /2\pi \Gamma )^{-1/3}N^{-1}
[\rho (1-\rho )]^{2/3}$ for the array with $N=80$ and different $M$.
Even at this value of $N$ the plateau corresponding to 
Eq.\ (\ref{S(w)}) is not yet very wide. 
With decreasing $M$ the plateau shrinks and there is a noticeable deviation 
from Eq.\ (\ref{S(w)}). Nevertheless, the numerical results presented 
in Fig.\ \ref{circle2} generally confirm the analytical result. 

        Comparing Eq.\ (\ref{S(w)}) with Eq.\ (\ref{largeomega}) it
is simple to estimate the frequency of the crossover to the
high-frequency limit: $\omega_{h}/2\pi \sim \Gamma [\rho (1-\rho )]^2$, 
which coincides with the frequency scale of ``collisions'' of an electron 
(or a hole) with its neighbors. Notice that for long arrays ($N\gg 1$) 
the high-frequency crossover shape does not depend on $N$ (similarly 
to the linear array case -- see Fig.\ \ref{n0.5}). 

        At low frequency Eq.\ (\ref{S(w)}) becomes invalid
when the size scale $L$ corresponding to the frequency $\omega_L$
becomes comparable with the total array length $N$. This allows us to 
estimate the position of the low-frequency crossover: $\omega_{l} /2\pi
\simeq {\tilde C} \Gamma [\rho (1-\rho )]^{1/2} N^{-3/2}$, where ${\tilde C}$
is a numerical factor. So, we have explained the dependence $\omega_{l}
\propto N^{-3/2}$ seen in Fig.\ \ref{circle0.3}. One can also check 
that at this frequency the result given by Eq.\ (\ref{S(w)}) 
transforms into Eq.\ (\ref{Fcircle}).

        It is interesting to find out at which electron concentration the
single-particle result $F=1/N$ becomes invalid. For $N\gg 1$ and small
number of electrons \cite{Derrida} $F\simeq (M!)^{2}2^{2M-1}/(2M)!N$, so
that considerable deviation from the single-particle result starts already
from $M=2$ and scales as $M^{1/2}$. This reflects the fact that in 1D arrays, 
significant correlation of hops starts at very small concentrations because 
randomly drifting electrons cannot pass each other.

\subsection{Temperature effect}

In the case of finite temperature when $\Gamma ^{-}\sim \Gamma ^{+}$,
the population of all charge configurations remain equal, so the 
average currents satisfy equation 
        \begin{equation}
\langle I^{\pm }\rangle =e\Gamma ^{\pm }\rho (1-\rho ),  
        \label{IT}\end{equation}
and the net current $\langle I\rangle =\langle I^{+}\rangle -\langle
I^{-}\rangle $ is still given by Eq.\ (\ref{<I>}) with $\Gamma =\Gamma^+
-\Gamma^-$. Plugging it into Eq.\ (\ref{Sinf}), we get 
        \begin{equation}
S_{I}(\infty )/2e\langle I\rangle =\frac{1}{N}\coth (W/2T).
        \label{STinf}\end{equation}
This result formally coincides with Eq.\ (\ref{Ssmall}), but now it is 
only valid for sufficiently high frequencies.

        Figure \ref{Temp}a shows the result of the Monte-Carlo simulations 
for the frequency dependence of the current spectral density. 
As the temperature 
$T$ is raised beyond the energy difference $W$, thermal fluctuations
gradually overwhelm the correlation effects, so that the high-frequency
plateau described by Eq.\ (\ref{STinf}) raises and gradually ``floods''
regions of lower and lower frequencies. [The fact that the low frequency part 
of the curve is less affected by thermal fluctuations can be interpreted as 
follows. 
Our arguments for Eqs.\ (\ref{S(w)}) and (\ref{Fcircle}) were based 
only on equal distribution of states and Eq.\ (\ref{<I>}) for the 
average current, which both remain unchanged for arbitrary temperature.
So, as long as the temperature is small enough so that Eq.\ (\ref{<I>}) 
is still applicable for the analysis of fluctuations at the frequency of
interest, the result is virtually unchanged.] 
$S_{I}(\omega )$ may be approximately found as the largest of values given by 
Eq.\ (\ref{STinf}) and the zero temperature result. 
As soon as $T \gtrsim 
T_{c}=W[N\rho (1-\rho )]^{1/2},$ the fluctuations are essentially thermal at
all frequencies, and the Fano factor is given by the Nyquist expression 
        \begin{equation}
F=\frac{2T}{NW}.
        \end{equation} 
Notice that as in the single-particle approximation, at $N\gg 1$ there is
a broad temperature region ($W N^{1/2}\ll T\ll WN$) where the array
is in thermal equilibrium, while the Fano factor is still much less than 1.

\subsection{Coulomb interaction effects}

        Coulomb interaction reduces the concentration fluctuations,
so one could also expect a decrease of the current fluctuations.  
This is illustrated in Fig.\ \ref{Coulomb1} which
shows typical Monte-Carlo results for zero temperature. 
One can see that as soon as $e^{2}/a$ becomes comparable or 
larger than $W$, the low-frequency fluctuations are gradually suppressed 
and can closely approach the limit (\ref{F=1/N}). Figure  \ref{Coulomb2} 
shows a typical dependence of the Fano factor on the array length $N$ for 
moderate values of the relative Coulomb interaction strength 
$r\equiv e^{2}/Wa$. 
At relatively small $N$ the scaling $F(N)$ is in between $N^{-1}$ and 
$N^{-1/2}$, while eventually at large $N$ it reaches the dependence 
$F\propto N^{-1/2}$ similar to the case without
Coulomb interaction. The presence of this transition  
is specific for 1D case, since in 1D systems the Coulomb interaction
cannot provide long-range electroneutrality (because the electric field 
$(\rho L e)/L^2$ produced by a charged fragment of length $L$, decreases 
with $L$). Hence, at large scale the density fluctuations are 
Coulomb-decoupled, which makes the general idea of the Fano 
factor derivation in Subsection C valid, leading
to the scaling $F\propto N^{-1/2}$. (In contrast, in 3D case the Coulomb
interaction does provide effective long-range electroneutrality, so $F$ 
inversely proportional to the system size is expected.) 

Stronger Coulomb interaction ($r \gtrsim [\min (\rho ,1-\rho )]^{-3}/2$) 
tries to fix the distance between 
the neighboring electrons and to turn them into a 1D Wigner crystal 
which may be rotated by the external field $W$. The Fano factor behavior 
in this case may be rather complex, because it depends on 
whether the integers $M$ and $N$ are ``commensurate'' (more strictly, 
whether their greatest common divisor is larger than 1) - 
see Fig.\ \ref{Coulomb3}. If it is, beyond some critical value $r_{c}$ of 
the ratio (about 2.6 for $N$ = 20 and $M$ = 10, see 
Fig.\ \ref{Coulomb3}) the Wigner crystal is stalled (at $T=0$), the system 
essentially turning into a Mott dielectric. At $r$ a little less
than $r_{c}$ the Fano factor starts to increase rapidly 
from $F\gtrsim 1/N$ to some value $F_{c}$; above $r_{c}$ the 
ratio $F=S_{I}(0)/2e\langle I\rangle$ is undetermined, since at $T$ = 0 there 
are neither fluctuations nor current. 
 In the opposite case of ``incommensurate'' $M$ and $N$ (the g.c.d. of $M$ and 
$N$ is 1) the Mott transition may be absent at $T=0$ even for arbitrary 
large $r$, and both the current and Fano factor may tend to the single 
particle results (\ref{single1}) and (\ref{Ssmall}), respectively. 
It is curious that on the way to this limit the function $F(r)$ may make 
a bump as if it tried to mimic the behavior of 
its commensurate counterpart - see Fig.\ \ref{Coulomb3}.

\section{Linear array}

\subsection{The model}

The main change associated with the linear array with external electrodes 
(Fig.\ \ref{schematic}a) is that the number $M$ of particles in  
the array is not more fixed. 
Instead, what is fixed are the chemical potentials 
of the metallic electrodes $\mu_{L,R}$ relative to the localized state 
energy. A model of the linear array should use this condition to specify 
rates of electron hopping between the electrodes and the edge localized 
sites. A reasonable way to reduce the number of additional parameters is 
to introduce two extra ``edge'' sites ($i=0,N$, not shown in Fig.\ 
\ref{schematic}a) which are very close to the electrodes. Then 
the ``edge'' tunneling rates $\Gamma _{L,R}^{\pm }$ are much higher than 
the ``bulk'' rates $\Gamma_i ^{\pm }$, so that the edge sites are in 
thermal equilibrium with the electrodes, and the probability of their 
occupation may be considered fractional but fixed: 
$f_{L,R}=[1+\exp (-\mu_{L,R}/T)]^{-1}$. In this approximation,
for a uniform array the rates of tunneling between the edge sites and 
their neighbors ($i=1$ and $i=N-1$) are related to the bulk rates 
$\Gamma^\pm$ as follows: 
\begin{eqnarray}
\Gamma _{1}^{+} &=&f_{L}\Gamma ^{+},\text{ \ }\Gamma
_{N}^{+}=(1-f_{R})\Gamma ^{+},  \label{edge_rates} \\
\Gamma _{1}^{-} &=&(1-f_{L})\Gamma ^{-},\text{ \ }\Gamma
_{N}^{-}=f_{R}\Gamma ^{+}.
\end{eqnarray}
We will be interested in the case of identical localized sites and similar
electrodes, so that $\mu _{L}=\mu _{R}$ and $f_{L}=f_{R}=f.$ 

	External electrodes also modify the Coulomb interaction of electrons. 
Besides that, the image charge effect 
makes the self-energy of the sites dependent on their location,
leading to nonuniform transport conditions.
Since in the present paper we concentrate on uniform arrays,
we will limit ourselves to the case of negligible Coulomb interaction.

\subsection{Global electron number fluctuation effects}

For the case of $T=0$, our model is reduced to the ASEP model 
with open boundaries. \cite{Derrida} Transport
properties for the latter model have been studied in detail, especially 
for $f_{L}=f_{R}=f$. In this case, the probability of any charge 
configuration is the same \cite{Derrida} as if each site had independent
occupation with probability $f$. As a consequence, the dc current
is given by Eq.\ (\ref{<I>}) with $\rho =f.$ The Fano factor can 
also be calculated analytically: \cite{Derrida2} 
	\begin{equation}
F=1-2f(1-f)\sum_{k=0}^{N-2}\frac{(2k)!}{k!(k+1)!}\,[f(1-f)]^{k} , 
	\end{equation}
and for $N\rightarrow \infty $ one finds a simple result:\cite{Derrida}  
	\begin{equation}
F=|1-2f|  ,
	\label{Fopen}\end{equation}
showing that the shot noise is much higher than in the circular arrays -- 
cf.\ Eq.\ (\ref{Fcircle}). Only in the evidently special point $f=1/2$, 
the Fano factor scales as in the closed boundary case \cite{Derrida2}: 
	\begin{equation}
F = (\pi N)^{-1/2}, \,\,\, N \gg 1.  
	\label{Fn0.5}\end{equation}

	Figure \ref{n0.3} (for $f=0.3$ and several values of $N$) 
shows that $S_{I}(\omega )$ smoothly decreases with frequency from
the value given by Eq.\ (\ref{Fopen}) and eventually reaches the level 
$S_{I}(\infty )=2eI/N$, in accordance with the general Eq.\ (\ref{Sinf}). 
As we will see later, at large $N$ the frequency dependence is quite
rich and exhibits three crossovers.

	The fact that the low-frequency shot noise in the linear array 
(Fig.\ \ref{schematic}a) at $f\neq 1/2$ is much higher than in the ring array 
(Fig.\ \ref{schematic}b) has a simple explanation: the total number of 
electrons in the case of ``open boundary conditions'' may significantly 
fluctuate, while on the ring this number is fixed. 
	Analytically, this effect may be especially simply considered 
for the case $f\ll 1$ (or similarly $1-f\ll 1$). 
Then the array is empty most of 
the time, and is entered very rarely by an electron (or 
hole). After the entry, the electron is transferred in a succession of hops
through the array, in total transferring charge $e$ from one electrode 
to another. This is exactly the situation for which the original Schottky
formula was derived, so that we get $F=1$ in agreement with the
corresponding limit of Eq.\ (\ref{Fopen}).

	In the case $f\ll N^{-1/2}$ (when electrons do not collide with
each other) 
the frequency dependence of $S_{I}$ can be obtained 
from Eq.\ (14) of Ref.\ \cite{Kor-set-an}, which was derived from the 
orthodox theory of single-electron tunneling for the similar sequential 
transport scenario. Assuming that 
tunneling rates are equal, $\Gamma _{i}^{+}=\Gamma $, besides the
negligibly small rate $\Gamma _{1}^+$, we get 
	\begin{equation}
\frac{S_{I}(\omega )}{2eI}=\frac{1}{N}+\frac{2}{N^{2}}\,\frac{\Gamma 
^{2}}{\omega ^{2}}\left[ 1-
\frac{ {\mbox R}{\mbox e} \,  
(1-\imath \omega /\Gamma )^{N-1}}{(1+\omega ^{2}/\Gamma ^{2})^{N-1}}\right] ;
	\end{equation}
we have confirmed this result using Monte-Carlo simulations. For $N\gg 1$
this formula is reduced to 
	\begin{equation}
\frac{S_{I}(\omega )}{2eI}=\left[ \frac{  
\sin (N\omega /2\Gamma )}{N\omega /2\Gamma } \right] ^2 ,
	\label{sinw/w}\end{equation}
that is obviously the normalized and squared Fourier image of the 
rectangular envelope of the train of $N$ current pulses during the single
electron passage.

 	For $f\sim 1$, Eq.\ (\ref{Fopen}) may be interpreted as follows. 
Let us again apply Eq.\ (\ref{<I>}) to long-range fluctuations. Then since 
$\partial I/\partial
\rho =e\Gamma (1-2\rho )$, one finds $S_{I}(0)=e^{2}\Gamma ^{2}(1-2\rho 
)^{2}S_{\rho }(0)$, where $S_{\rho }(0)$ is the low-frequency
intensity of fluctuations of the total array occupation $(\rho \equiv M/N)$.
Notice that for $\rho =f =1/2$ the result vanishes, and we should go after the
higher order effect as we did for the ring array. 
For all other values of $f$, we may use the estimate $S_{\rho }(0) \sim 
\langle (\delta \rho ) ^{2}\rangle /\Delta \omega $, where
$\langle (\delta \rho ) ^{2}\rangle =f(1-f)/N$ and  the
effective bandwidth $\Delta \omega $ can be estimated as $|v_{s}/N|$ 
(unlike in the ring array, the density fluctuation
is carried out of the linear array with velocity $v_{s}$ given by Eq.\ 
(\ref{vs})). 
Combining these formulas, we obtain  
$\Delta \omega \sim \left| 1-2f\right| \Gamma /N,$ $S_{\rho }(0)\sim
f(1-f)/\Gamma |1-2f|$, and $F=const\times |1-2f|$. The numerical factor 
in this result for the Fano factor cannot be derived in this crude way, 
but it obviously equals 
unity because at $f=1$ we should get the previous result, $F=1.$ Thus we 
completely recover the exact result (\ref{Fopen}).

	One more possible derivation of that equation can be obtained along 
the following line. If $f<1/2$, then the electrons can be supplied from 
the left electrode with the
maximum rate $f\Gamma $, while the average ``sink'' velocity $(1-f)\Gamma $
is larger. Hence, only electrons relatively close to the left boundary can
affect the entrance of the next electrons, and so the low-frequency 
correlation is essentially the boundary effect. Using this idea and taking 
into account, for example, correlations only due to the three first jumps, 
it is easy to obtain $F=1-2f+{\cal O}(f^{3})$. Taking into account 
more jumps we would eventually show that Eq.\ (\ref{Fopen}) is exact. 

	A concentration fluctuation supplied from the boundary moves 
with velocity $v_s$, so for $N\gg 1$ the corresponding envelope in 
$I(t)$ has rectangular shape with duration $N/|v_s|$. 
Combining the corresponding frequency dependence of $S_I(\omega )$ with
the exact result for $F$, we get 
	\begin{equation}
\frac{S_{I}(\omega )}{2eI} \simeq |1-2f| \left[ \frac{  
\sin (N\omega /2\Gamma |1-2f| )}{N\omega /2\Gamma |1-2f| } \right] ^2 . 
	\label{S(w)open}\end{equation}

	At sufficiently high frequency, $\omega \gg |v_s|/N$, 
the ``nonlinear'' contribution from the concentration fluctuations 
obviously should be the same in the linear and ring arrays (with equal 
average concentration $\rho =f$). Hence, $S_I(\omega )$ will
still be given by Eq.\ (\ref{S(w)}) while in the crossover region 
it can be crudely estimated as a sum (or maximum value) of 
two contributions given by Eqs.\ (\ref{S(w)open}) and (\ref{S(w)}).
As a result, there are three characteristic frequencies in $S_I(\omega )$
dependence at $N\gg 1$: the low-frequency saturation occurs at 
$\omega \lesssim \omega_{l}\sim \Gamma |1-2f|/N$, the 
intermediate-frequency
dependence described by Eq.\ (\ref{S(w)}) starts at $\omega \gtrsim
\omega_{m} \sim \Gamma N^{-3/5} |1-2f|^{9/5} [f(1-f)]^{-2/5}$,
and finally the high frequency saturation occurs at $\omega \gtrsim
\omega_{h} \sim \Gamma [f(1-f)]^2$, similar to the ring array case. 

	The case $f=1/2$ plays a special role in the ASEP theory, 
as can be easily noticed comparing Eqs.\ (\ref{Fopen}) and (\ref{Fn0.5}).
Actually, this case is quite important since for sufficiently long arrays 
with $f_{L}>1/2$
and $f_{R}<1/2$ the electron concentration in the bulk of the array is close 
\cite{Derrida} to $f=1/2$ (so $\langle I\rangle = e\Gamma /4$) 
and, hence, the scaling 
$F \propto N^{-1/2}$ holds as in Eq.\ (\ref{Fn0.5}).
 (As an example, for $f_{L}=1$, $f_{R}=0$ the result is \cite{Derrida2} 
$F = 3(2\pi )^{1/2}/16N^{1/2}$). 
	At $f=1/2$ the low-frequency fluctuations can no longer be 
considered as a boundary effect, since the ``sink'' velocity $(1-f)\Gamma $ 
is equal in this case to the maximum supply rate $f\Gamma $; hence, 
the transport becomes jammed and the correlations involve the
whole array length. 

	In the respect that the boundary effects are no longer important,
the linear array at $f=1/2$ is very similar to the circle array. 
Figure \ref{n0.5} shows the frequency dependence of the current 
spectral density for $f=1/2$ and several values of $N$. The data  
look similar to that in Fig.\ \ref{circle0.3}.
The main feature is $\omega^{-1/3}$ dependence in the intermediate 
frequency range. To check the validity of Eq.\ (\ref{S(w)}) in
this range, Fig.\ \ref{n0.5}b shows the same data as Fig.\ \ref{n0.5}a
but normalized by $S_0(\omega )= 2eI (\omega /2\pi )^{-1/3}N^{-1}
[f(1-f)]^{2/3}$. We see that as $N$ grows, the intermediate region 
becomes more and more pronounced.

\subsection{Temperature effects}

Figure \ref{Temp-lin} shows the numerically calculated effect of 
nonvanishing temperature on the shot noise in a linear array. It shows 
that the effect is quite similar to that in a ring array 
(Fig.\ \ref{Temp}), however, because of the higher 
initial intensity of low-frequency fluctuations (at $T=0$) the noise becomes
completely thermal at a higher temperature, $T\gtrsim W|1-2f|$.

\section{Discussion} 

	Probably the most important result of our analysis is that 
in contrast to the expectation based on the analysis of 1D arrays of 
conventional tunnel junctions, the shot noise in 
uniform 1D hopping arrays is typically much higher than $1/N$ of the
Schottky value $S_I=2e\langle I\rangle$. 
However, in some cases this lower bound can be achieved. In order 
to sort out these cases, it is useful to consider the current
fluctuations as the result of three major sources:

- time randomness of electron tunneling events,
 
- electron density fluctuations, and

- thermal fluctuations. \\
Crudely speaking, the lower bound $2e\langle I\rangle/N$ for the noise 
is determined by the first contribution, while the second contribution 
typically increases the noise significantly even at $T=0$. 

	At relatively high 
frequencies the current spectral density in a ring array and 
1D array between electrodes behaves pretty similarly. In particular, 
the high-frequency asymptote is given by the same Eq.\ (\ref{Sinf}) 
and is determined by capacitive factors $\lambda_i$. (This result
is also valid for the conventional case of 1D array of tunnel junctions
 -- see Appendix). If $\lambda_i=1/N$, then in all cases 
at low temperature $T$ we have $S_I(\infty )=2e\langle I\rangle /N$.

 	However, at low frequency the noise behavior in a ring array and 
a linear array is quite different. The reason for the difference 
is that in a linear array the total number of electrons can
fluctuate while in the ring array it is fixed. 
In the case when the single-particle approximation is applicable for
a linear array, the relative density fluctuations are maximal, and
the Fano factor is not suppressed: $F=1$ at $T=0$. 
The electron ``collisions'' (the Pauli exclusion) reduce these 
fluctuations, but quite inefficiently. 
Only in the special case of half filling ($f=0.5$) when ``traffic jams'' 
have all 
size scales, the Fano factor decreases as $N^{-1/2}$ with the array 
length $N$; in other cases the dependence $F(N)$ quickly saturates at
the level $F=|1-2f|$. 
One can speculate that Coulomb interaction should be a more efficient 
factor in suppression of $F$, since it may significantly reduce the 
electron density fluctuations, however, this conclusion has still 
to be verified numerically.  

	In contrast to the linear array, in the uniform ring array the 
uncorrelated motion (of a single electron) provides the maximal 
suppression of the Fano factor,
$F=1/N$. The Pauli exclusion in fact increases $F$ leading
to $F\propto N^{-1/2}$. However, the extra correlations due to
Coulomb interaction between electrons on different sites make
transport ``smoother'' and reduce the Fano factor, in some cases
down to the lower bound $F=1/N$. 

	It is instructive to compare these results with those for a 1D array 
of tunnel junctions (see Appendix). In the latter model the Fano factor 
is determined purely by the junction resistances. In some sense, 
this is a consequence of strong Coulomb interaction which forbids noticeable 
charge fluctuations and establishes fast long-range correlations 
between currents through different tunnel junctions. In the uniform array 
at low temperature the noise 
suppression is maximal, $F=1/N$. However, if the junctions are very small, 
single-electron effects can lead to significant charge 
fluctuations and thus increase the Fano factor. For example, $F=1$ 
is realized \cite{Matsuoka} in the vicinity of the Coulomb blockade 
threshold when the transport has a bottleneck even in the uniform array. 

	So far we have reviewed our results for the uniform case. Now 
let us briefly 
discuss hopping transport noise in nonuniform 1D arrays. It is simple 
to study one particle inside the ring array 
with arbitrary disorder at low temperatures. In this case the transport 
is unidirectional, $\Gamma_i^-=0$, and the average current is obviously 
given by the expression 
	\begin{equation}
\langle I\rangle=e \left[ \sum_i (\Gamma_i^+)^{-1}\right] ^{-1} ,  
	\label{Idis}\end{equation}
while the formula for the current spectral density has been derived in Ref.\ 
\cite{Kor-set-an} for the case $\lambda_i=1/N$, and can be readily 
generalized to include arbitrary $\lambda_i$: 
	\begin{eqnarray}
&& S_{I}(\omega ) = 2e\langle I\rangle \sum_{l=1}^N \lambda_l^2 + 
4e\langle I\rangle \, \mbox{Re} \left\{ \left[
\prod_{l=1}^N \left( 1+ \frac{\imath \omega}{\Gamma_l^+} \right) -1\right]
^{-1} \right.  \nonumber \\
&& \left. \times \left[ \sum_{l=1}^N \lambda_l^2 + 
\sum_{l=1}^{N-1} \sum_{m=1}^N \lambda_m \lambda_{m+l} 
\prod_{k=1}^l \left( 1+\frac{\imath \omega}{\Gamma_{k+m}^+}
\right) \right] \right\} ,  \label{S(w)dis}
\end{eqnarray}
where by definition $\Gamma_{N+k}^+=\Gamma_{k}^+$. At zero frequency this
formula is reduced to 
	\begin{equation} 
\frac{S_{I}(0)}{2e\langle I\rangle } =\left[ \sum_i (\Gamma_i^+)^{-2}\right] 
\left[ \sum_i (\Gamma_i^+)^{-1} \right] ^{-2},  
	\label{Fanodis}\end{equation}
and allows study of the statistics of the Fano factor for random 
distribution of $\Gamma_i^+$ in a long array, $N\gg 1$. 

As the major factor,
let us take into account the dependence of tunneling rate on the distance 
$a_i$ between sites, $\Gamma_i^+=\Gamma_0 \exp (-2a_i/\xi )$, where $\xi$ 
is the localization length, and assume that independent random $a_i$ obey the 
Poisson distribution, $p(a_i)=a_0^{-1}\exp (-a_i/a_0)$ where $a_0 \gg \xi$ is 
the average spacing. Then the distribution of rates can be parameterized as 
$\Gamma_i^+=\Gamma_0 \, x_i^{2a_0/\xi }$, where the random number $x_i$ has 
uniform distribution between 0 and 1. The minimal rate (``bottleneck'') 
$\Gamma_{min}$ will be about $\Gamma_0 (2a_0/\xi\mbox{e}N)^{2a_0/\xi}$ on 
average (here $\mbox{e}=2.71..$), while the next minimal rate 
$\Gamma_{min+1}$ will be much larger, $\Gamma_{min+1}/\Gamma_{min} 
\sim (2a_0/\xi \mbox{e}N)^{-2a_0/\xi} 
\gg 1$. It is easy to see that in this case both the average current [Eq.\ 
(\ref{Idis})] and the Fano factor [Eq.\ (\ref{Fanodis})] are determined by
the bottleneck: $\langle I\rangle =e\Gamma_{min}$ and $F=1$. 

It is also instructive to consider a model where the maximal distance 
$a_{i}$ is limited by some big value $a_{max}$ ($a_{max}\gg a_{0}$). For 
example, this describes the situation in which 
some other transport mechanism starts to dominate over tunneling 
when the sites are too far apart, thus 
limiting $\Gamma $ from below. If $N\ll \exp (a_{max}/a_{0})$, the results
for the average current and the Fano factor do not differ from the case
considered above. However, for very long arrays, $N\gg \exp (a_{max}/a_{0})$,
the transport is limited by many similar bottlenecks, so that 
$\langle I\rangle \simeq
e\Gamma _{0}\exp (-2a_{max}/\xi +a_{max}/a_{0})/N$ and the Fano factor 
decreases with the array length, $F\simeq \exp (a_{max}/a_{0})/N$. 

	We can use this result for a preliminary estimate of the Fano 
factor for hopping in disordered 2D and 3D systems when the transport is 
mainly determined by percolation clusters. \cite{Shklovskii} 
	If the single-particle approach is applicable, then as in 
the case above we can simply count the number of similar bottlenecks 
in the transport direction. With this 
argumentation, we obtain a simple estimate
\begin{equation}
F \sim L_c/L,  
	\label{cluster}\end{equation}
where $L$ is the sample length and $L_c$ is the characteristic size of 
the percolation cluster. The applicability range of the approach
is rather unclear, so this result still has to be confirmed using 
either more quantitative analysis or numerical Monte-Carlo modeling.
(Preliminary numerical analysis of hopping in uniform 2D arrays gives 
an indication of the dependence $F\propto L^{-\nu }$ where $\nu \sim 0.8$ 
is rather close to unity.)

	In conclusion, in the present paper we have studied the shot
noise at hopping in 1D arrays of sites, concentrating on the 
uniform case and briefly considering the effect of disorder. 
It is important to extend this study to 2D and 3D hopping. 
The presented results hint that the Coulomb interaction may 
play the crucial role in the suppression of low-frequency shot noise.

\section{Acknowledgment} 
Fruitful discussions with D.\ A.\ Parshin and V.\ V.\ Kuznetsov are 
gratefully acknowledged. 
This work was supported in part by the Engineering Research Program of 
the Office of Basic Energy Sciences at the Department of Energy. 

\appendix

\section{1D array of tunnel junctions} 

In this Appendix we calculate the spectral density of the shot noise in
a 1D array of tunnel junctions. We will first use the standard circuit theory 
and then extend the calculations to the case of weak single-electron effects
(which are assumed to be small because of high temperature or high current).

Figure \ref{junctions} shows the array of $N$ tunnel junctions in series.
Each junction is characterized by resistance $R_{i}$ and capacitance $C_{i}$
($i=1,\ldots N$), while $C_{g,k}$ ($k=1,\ldots N-1$) denote capacitances of
``islands'' to the ground. This circuit does not describe the long-range
capacitances which can be especially important for islands and junctions 
of small size, \cite{Likh-Matsuoka} that can require numerical calculation of
the total capacitance matrix. \cite{Likh-Matsuoka,Kor-wireless,Kor-param}
The derivation below is valid for arbitrary capacitance matrix, however, for
simplicity we will refer to Fig.\ \ref{junctions}.

In the absence of single-electron correlations the {\it I-V} curve of 
the array is linear, $\langle I\rangle =V/R_\Sigma$, $R_\Sigma =\sum_i R_i$, 
and the average voltage across each junction is proportional to the 
junction resistance, $\langle V_i\rangle =\langle I\rangle R_i$. 
To study the fluctuations 
we follow the standard circuit theory \cite{Ziel} and 
introduce the sources of the current noise $\xi_i(t)$ in parallel with the
junctions. Separating the current through 
$i $th junction into two parts flowing in opposite directions, 
\begin{equation}
\langle I\rangle =\langle I_i^+\rangle  -\langle I_i^-\rangle , \,\,\,
 \langle I_i^+\rangle /\langle I_i^- \rangle = \exp (-e\langle V_i\rangle /T), 
\end{equation}
and using the Schottky formula for each part, we get the following white
spectral density for the noise source $\xi_i(t)$: 
\begin{equation}
S_{\xi i}(\omega )=2e\langle I_i^+\rangle  +2e\langle I_i^-\rangle 
=2e\langle I \rangle \, \mbox{coth} (e\langle V_i\rangle /2T).
\label{Sxi}
\end{equation}

Let us denote by $\phi_i(t)$ the fluctuating part of the $i$th island
potential ($\phi_0(t)=\phi_N(t)=0$ because we assume constant potentials of
the leads), then the current $I_i(t)$ through $i$th junction can be written
as 
	\begin{equation} 
I_i(t) = \langle I\rangle +[\phi_{i-1}(t) -\phi_i(t)]/R_i +\xi_i(t).  
	\label{I_i}\end{equation}
The evolution of $\phi_i (t)$ is described by the equation 
\begin{equation}
\dot{\phi}_i = \sum_{j=1}^N I_k (t)  \left[ D_{i,j} - D_{i,j-1} 
\right] , 
\end{equation}
where ${\bf D} \equiv {\bf C}^{-1}$ is the inverse capacitance matrix 
[obviously $D_{i,j}=D_{j,i}$ and $D_{0,i}=D_{N,i}=0$] and can be 
rewritten in the following form: 
\begin{eqnarray}
\dot{\phi}_i &=& \sum_{k=1}^{N-1} A_{i,k} \phi_k + \sum_{k=1}^{N} B_{i,k}
\xi_k ,  \label{phi} \\
A_{i,k} &\equiv & B_{i,k+1}/R_{k+1} - B_{i,k}/R_k, \\
B_{i,k} &\equiv & D_{i,k} -D_{i,k-1}.
\end{eqnarray}
Notice that ${\bf A}$ is $(N-1)\times (N-1)$ matrix while ${\bf B}$ is 
$(N-1)\times N$ matrix. In the frequency representation Eq.\ (\ref{phi}) can
be written as $\imath \omega \, {\bf \phi} (\omega ) = {\bf A \, \phi}
(\omega ) +{\bf B \, \xi} (\omega)$ ($\imath$ is the imaginary unity) and
can be easily solved in the matrix form, 
\begin{equation}
{\bf \phi} (\omega ) =\left( \frac{{\bf 1}} {\imath\omega {\bf 1} -{\bf A}}
\right) {\bf B \, \xi }(\omega ).
\end{equation}
Using Eq.\ (\ref{I_i}) we find the Fourier transform of $I_i (t)$: 
\begin{eqnarray}
I_i(\omega )= && \sum_{j=1}^N X_{i,j} (\omega ) \, \xi_j(\omega ),
\label{Iomega} \\
X_{i,j}(\omega )= && \delta_{ij} + \frac{1}{R_i} \sum_{k=1}^{N-1} \left[
\left( \frac{1}{\imath\omega -{\bf A}}\right)_{i-1,k} \right.  \nonumber \\
&& \left. - \left( \frac{1}{\imath\omega -{\bf A}}\right) _{i,k} \right]
B_{k,j} \, ,  \label{Xij}
\end{eqnarray}
where by definition $(\imath\omega -{\bf A})^{-1}_{0,k}= (\imath\omega -
{\bf A})^{-1}_{N,k}=0$. 

Notice that at $\omega \rightarrow \infty $ the only surviving term in Eq.\ 
(\ref{Xij}) is the Kronecker symbol $\delta_{ij}$ so that 
\begin{equation}
I_i(\infty )=\xi_i (\infty ).  \label{winf}
\end{equation}
At $\omega =0$ it is possible to prove (using somewhat cumbersome algebra)
the relation 
\begin{equation}
I_i(0)=\sum_j (R_j/R_\Sigma) \, \xi_j(0),  \label{wzero}
\end{equation}
which obviously means that at low frequencies the current is distributed
according to resistances and equal in all junctions. (Eq.\ (\ref{wzero})
shows that the fraction $R_j/R_\Sigma$ of the current $\xi_j$ flows through
the array while the rest is returned via the ``shunt'' $R_j$).

The spectral density of the current $I_i$ through $i$th junction can be
readily calculated as 
\begin{equation}
S_{Ii}(\omega )= \sum_j |X_{i,j}(\omega )|^2 S_{\xi j},  \label{SIi}
\end{equation}
because the noise sources $\xi_i$ are mutually uncorrelated. Using Eqs.\ 
(\ref{Sxi}), (\ref{winf}) and, (\ref{wzero}) we get the simple expressions in
the limiting cases: 
\begin{eqnarray}
&& S_{Ii}(0) = 2e\langle I\rangle \sum_j (R_j^2/R_\Sigma^2) \, 
\mbox{coth}(e\langle I\rangle R_j/2T) ,
\label{SI0} \\
&& S_{Ii}(\infty ) = 2e\langle I\rangle \, 
\mbox{coth} (e\langle I\rangle R_i/2T) .
\end{eqnarray}

In experiment it is usually impossible to measure the current through 
one junction, and the only measurable quantity is the current in the
external lead which contains the contribution from the displacement current
and, hence, depends on the currents through all junctions. The current 
$I(t) $ at the left external lead can be expressed as the linear
combination, 
\begin{eqnarray}
&& I(t) = \sum_i \lambda_i I_i(t),  \label{Ie} \\
&& \lambda_i = \delta_{1i} +\sum_j C_{0,j} B_{j,i}, \,\,\,\, 
\sum_i \lambda_i =1,
\label{Ki}
\end{eqnarray}
where $C_{0,k}$ is the element (always negative) of the capacitance matrix
between the left electrode and $k$th island. In the case of Fig.\ 
\ref{junctions},
$C_{0,k}=-C_1 \delta_{1k}$ because the left electrode is capacitively
coupled only with the first island, so $\lambda_i=\delta_{1i}-C_1 B_{1,i}$. 
If all $C_{g,i}=0$ in Fig.\ \ref{junctions}, 
then $\lambda_i=C_i^{-1}/\sum_jC_j^{-1}$,
while for a long uniform array, $C_i=C=const$, $C_{g,i}=C_g=const$, $N^2 \gg
C/C_g$, the coefficients are $\lambda_i={\cal X}^{i-1} (1-{\cal X})$ 
where ${\cal X}=1+C_g/2C- [(C_g/2C)^2+C_g/C]^{1/2}$. 
If the tunneling system is different from what is shown in Fig.\ 
\ref{junctions} and consists of small islands which are separated 
by much larger distances and imbedded into the plane capacitor (external 
electrodes), then similar to the hopping case 
$\lambda_i=a_i/\sum_j a_j$ where $a_i$ is the length of the projection of 
$i$th tunneling jump onto the direction perpendicular to the capacitor planes.

Using Eqs.\ (\ref{Ie}) and (\ref{Iomega}) we obtain the following expression
for the spectral density of the left external current:
\begin{equation}
S_{I}(\omega )= \sum_{j} \left| \,  \sum_i \lambda_i X_{i,j}(\omega ) 
\right| ^2 S_{\xi j} 
\label{SIe}
\end{equation}
(notice that at finite frequency, $S_I(\omega )$ for the left and right 
electrodes can be different). 
In the particular case shown in Fig.\ \ref{junctions},  
\begin{eqnarray}
\sum_i \lambda_i X_{i,j} = && \delta_{1j} - C_1 B_{1,j} - 
\sum_{m,k}^{N-1} \left[ 
\frac{\delta_{1m}}{R_1} +C_1 A_{1,m} \right]  \nonumber \\
&&\times \left( \frac{1}{\imath\omega -{\bf A}} \right) _{m,k} B_{k,j} .
\label{KiXij}
\end{eqnarray}

In the limit $\omega =0$ the spectral density $S_{I}(0)$ of the external
current coincides with the spectral density of the current through any
junction (see Section II) and, hence, is given by Eq.\ (\ref{SI0}). 
While at low frequency $S_{I}$ is 
determined only by the circuit resistances, in the opposite limit, $\omega
\rightarrow \infty$, it is determined mainly by the circuit capacitances, 
\begin{equation}
S_{I}(\infty ) =\sum_i \lambda_i^2 \, S_{\xi i} = 2e\langle I \rangle 
\sum_i \lambda_i^2 \, \mbox{coth} (e\langle I\rangle R_i/2T)
\end{equation}
(resistances are important only when the Nyquist noise contribution is
considerable, $IR_i \lesssim T$). It is simple to check that in the thermal
equilibrium, $I=0$, the low-frequency noise [see Eq.\ (\ref{SI0})] always
satisfies the Nyquist formula, $S_{I}(0)=4T/R_\Sigma$, however, this is not
true at finite frequency, for example $S_{I}(\infty )=4T\sum_i 
\lambda_i^2/R_i$.
Notice that in this formalism $\omega =\infty$ still means $\hbar \omega \ll
\max (T,eV_i)$. However, it would be simple to take into account zero-point
fluctuations by replacing Eq.\ (\ref{Sxi}) with \cite{Dahm} $S_{\xi 
i}(\omega )=R_i^{-1} \sum_\pm (eV_i\pm \hbar \omega) \, \mbox{coth}[(eV_i\pm
\hbar \omega )/2T]$.

In a uniform array at zero temperature the noise at low frequency is 
suppressed $N$ times [see Eq.\ (\ref{SI0})] in comparison with the Schottky
formula, $S_{I}(0)=S_{Ii}(0)=2eI/N$. However, at high frequency the
suppression of the external current noise is usually weaker, $S_{I}(\infty
)=2eI\sum_{i}\lambda_{i}^{2}$, $1/N\leq \sum_{i}\lambda_{i}^{2}\leq 1$, 
while for the
current through a particular junction there is no suppression at all, 
$S_{Ii}(\infty )=2eI$.

It is interesting to find out at which $\omega $ the low-frequency result is
no longer accurate. We have studied the long uniform arrays shown in Fig.\ 
\ref {junctions}, $N\gg (C/C_g)^{1/2}$, and found numerically that the 
relative 
accuracy $\epsilon \ll 1$ of the low-frequency result, $S_{I}(\omega )/2eI
=(1+\epsilon )/N$, corresponds to the frequency $\omega /2\pi \simeq 1.1
(RC_g)^{-1} \epsilon^{1/2}N^{-2}$. This formula, however, cannot be used to
describe the crossover to the high-frequency asymptote, $S_{I}(\infty 
)=(1+4C/C_g)^{-1/2}$.

\vspace{0.3cm}

If the
typical size of the tunnel junctions is small, single-electron effects 
\cite{AvLikh91} become important. Below we extend the standard noise theory
to this case, assuming that single-electron effects are weak due to
relatively high temperature, $T\gtrsim e^2/{\tilde C}$, or relatively high
current, $I\gtrsim e/{\tilde R}{\tilde C}$ ($\tilde C$ and $\tilde R \gg R_Q$
are the typical capacitance and resistance of tunnel junctions).

According to the ``orthodox'' theory, \cite{AvLikh91} the rate of tunneling
through $i$th junction (in the positive direction), 
\begin{equation}
\Gamma = V^{eff}_{i}/eR_i[1-\exp(-eV^{eff}_{i}/T)]  \label{Gamma}
\end{equation}
is governed by the effective voltage $V^{eff}_{i}$ which is always smaller
than the actual voltage $V_i$, 
\begin{eqnarray}
&& V^{eff}_{i}=V_i - e/2C_{t,i}, \\
&& 1/C_{t,i} = D_{i-1,i-1} + D_{i,i} -2 D_{i-1,i},
\end{eqnarray}
where $C_{t,i}$ is the total capacitance of $i$th junction. (For tunneling
in the opposite direction the effective voltage is $-V_i-e/2C_{t,i}$.)
Linearizing Eq.\ (\ref{Gamma}) and averaging over the fluctuating $V_i(t)$,
we obtain the equation 
\begin{eqnarray}
\langle I \rangle = && \langle I_i^+ \rangle - \langle I_i^- \rangle , \\
\langle I_i^\pm \rangle = && \frac{\pm \langle V_i\rangle -e/2C_{t,i}} 
{R_i [1-\exp (-e(\pm \langle V_i\rangle -e/2C_{t,i})/T)]} \, ,
\end{eqnarray}
which allows the calculation of average voltages $\langle V_i\rangle$ 
for the given average current $\langle I\rangle$. (Notice that the 
high-voltage offset of the {\it I-V} curve is equal to 
$V_{off}=\sum_i e/2C_{t,i}$ exactly.) This approximation has been 
successfully used for the analytical and numerical analysis of the {\it I-V}
curves of Coulomb blockade thermometers. \cite{Pekola} 

Since we assumed an essentially linear response of any junction current, it 
is natural to use the formalism of the standard circuit theory, so the result
for the current spectral density will be given by the same Eqs. (\ref{Xij}),
(\ref{SIi}), and (\ref{Ki})--(\ref{KiXij}). However, the second equality in
Eq.\ (\ref{Sxi}) is no longer valid, and we have at least three choices for
the ``seed'' noise spectral density: $S_{\xi i} =2e\langle I_i^+ \rangle 
+2 e\langle I_i^-\rangle$, $S_{\xi i} = 2e\langle I\rangle \mbox{coth}  
(e\langle V_i\rangle /2T)$, or $S_{\xi i} = 2e\langle I \rangle \mbox{coth} 
(e\langle I\rangle R_i/2T)$. The first choice seems to be the most natural 
one, however, 
numerical comparison with the results of Monte-Carlo simulations shows that
the first formula usually underestimates noise (see Fig.\ \ref{Sjunctions}),
the third formula overestimates it, and the second formula (which is in
between two others) usually gives the closer result, though not always.
Notice, however, that all three approximations coincide in the limits of
both low and high temperature, so the difference between them is never too
significant within the applicability range of the formalism.

As the next level of approximation for the ``seed'' noise $S_{\xi i}$ at
finite temperature, it is possible to estimate the standard deviation of
fluctuating $V_i(t)$ and take into account the correction due to the second
derivative of Eq.\ (\ref{Gamma}). It is also possible to take into account
the effective increase of the junction resistances used in the evolution
equation (\ref{I_i}) at finite temperatures (we did not implement this last
idea numerically). Figure \ref{Sjunctions} shows the frequency dependence of
the spectral density of the current in the (left) external electrode for the 
uniform array
of $N=10$ junctions with $C_g/C=0.1$ symmetrically biased by voltage $V=5
e/C $ at temperature $T=e^2/C$. The thick line shows the results of
Monte-Carlo simulations while the thin lines represent the calculations
using Eqs.\ (\ref{SIe})--(\ref{KiXij}). For the lowest and highest thin
lines, the first and third formulas for $S_{\xi i}$ discussed above have
been used. The thin line corresponding to the second formula is almost
indistinguishable from the thin line showing the result using the second
(quadratic) approximation of $S_{\xi i}$. As one can see, these lines are
quite close to the Monte-Carlo result. With an increase of current or 
temperature all thin lines become closer to each other, and the agreement
with the Monte-Carlo result becomes even better.

\begin{figure}[tbp]
\caption{(a) Linear array of $N-1$ localized sites connecting two electrodes 
(``open boundary conditions''). The 
electron transport is determined by the tunneling rates $\Gamma _{i}^{\pm }$. 
(b) Circular array
(``periodic boundary conditions'') with $N$ sites occupied by $M$ electrons. }
\label{schematic}
\end{figure}

\begin{figure}
\caption{The Fano factor $F$ as a function of energy difference per site $W$
        in a circular array occupied by one electron.}
\label{coth}
\end{figure}

\begin{figure}[tbp]
\caption{Frequency dependence of the spectral density 
$S_I(\omega )$ for uniform circular arrays at $T=0$  
for several values of array lengths $N$ and electron concentration 
$\rho=M/N$: (a) $\rho=0.3$,
(b) $\rho=0.5$.}
\label{circle0.3}
\end{figure}

\begin{figure}[tbp]
\caption{ Current spectral density for the array in a circle normalized by 
$S_0(\protect\omega )= 2eI (\protect\omega /2\protect\pi 
)^{-1/3}N^{-1}[\rho (1-\rho )]^{2/3}$ [see Eq.\ (\ref{S(w)})]. }
\label{circle2}
\end{figure}

\begin{figure}[tbp]
\caption{Current spectral density for the ring array with $N=80$ and $M=
40$ for several temperatures.}
\label{Temp}
\end{figure}

\begin{figure}[tbp]
\caption{Current spectral density in a ring array at zero temperature 
for several values of Coulomb interaction strength 
$r = e^2/aW$. } 
\label{Coulomb1}
\end{figure}

\begin{figure}[tbp]
\caption{Fano factor of ring arrays with a fixed electron
concentration ($M/N=0.5$) as
function of the array length $N$, for several values of Coulomb 
interaction strength. Lines are just guides for the eye. }
\label{Coulomb2}
\end{figure}

\begin{figure}[tbp]
\caption{The dependence of the Fano factor for the ring array on the 
strength $r$ of Coulomb interaction. }
\label{Coulomb3}
\end{figure}

\begin{figure}[tbp]
\caption{Frequency dependence of the spectral density $S_{I}
(\protect\omega )$ for uniform linear arrays 
with symmetric boundary conditions, $f_{L}=f_{R}=0.3$, at $T=0$.
}
\label{n0.3}
\end{figure}

\begin{figure}[tbp]
\caption{(a) -- Current spectral density for linear array with 
$f_{L}=f_{R}=0.5$ at $T=0$. Notice the 
dependence $S_{I}(\protect\omega )\sim \protect\omega ^{-1/3}$ in the
intermediate frequency range between the saturations at low frequency 
($F\sim N^{-1/2}$) and high frequency ($S_{I}(\infty )/2eI=1/N$). 
(b) -- The same data normalized by $S_0(\protect\omega )= 2eI 
(\protect\omega /2\protect\pi )^{-1/3}N^{-1}[f(1-f)]^{2/3}$. 
 }
\label{n0.5}
\end{figure}

\begin{figure}[tbp]
\caption{Frequency dependence of the current spectral density 
in the linear array with $N=10$ and $f_L=f_R=0.3$ 
at several temparatures. 
}
\label{Temp-lin}
\end{figure}

\begin{figure}[tbp]
\caption{1D array of N tunnel junctions 
with capacitances 
$C_i$ and resistances $R_i$. $C_{g,i}$ are the capacitances to the ground.
The Langevin noise sources are presented by random current generators 
$\protect\xi_i (t)$ parallel to the junctions. }
\label{junctions}
\end{figure}

\begin{figure}[tbp]
\caption{Normalized spectral density $S_{I}(\omega )$ 
for a uniform array of small tunnel 
junctions. Thick line shows the result of Monte-Carlo simulations while thin
lines 1--4 are calculated using Eqs.\ (\ref{SIe})--(\ref{KiXij}). For lines
1--3 the ``seed'' noise $S_{\protect\xi i}$ is calculated as 
$2e(\langle I_{i}^{+}\rangle -\langle I_{i}^{-}\rangle )$, 
$2e\langle I\rangle \,\mbox{coth}(e\langle V_{i}\rangle /2T)$, and 
$2e\langle I\rangle \, \mbox{coth}(e\langle I\rangle R_{i}/2T)$, 
respectively. For the line 4 (by coincidence, 
almost indistinguishable from line 3) $S_{\protect\xi i}$ is calculated
using the quadratic approximation. }
\label{Sjunctions}
\end{figure}


\begin{references}
\bibitem{shot} There have been, however, several noise analyses
	\protect\cite{Derrida,Derrida2,Jensen} which are partly relevant 
	for our workgoals; we will refer to them at appropriate occasions
	below.
\bibitem{holes}  The results are also valid at $M=N-1$, when
only one hole is present in the array. Similar results would be also valid 
in the case of several electrons, if they were not correlated either due to 
the Pauli principle (imbedded into our model) or Coulomb interaction.
\bibitem{SOC} The self-organized criticality has been shown 
	\protect\cite{Jensen} to lead to $1/f$ power spectrum in 
	a somewhat different model 
	describing the flux flow in superconductors.
\end{references}
\end{document}